\documentclass[prb,aps,showpacs,unsortedaddress,superscriptaddress,amsmath,amssymb,floatfix,twocolumn]{revtex4}
\usepackage{graphicx}
\usepackage{amssymb}

\begin{document}

\title{Resonant inelastic X-ray scattering in a Mott insulator}
\author{Nandan Pakhira}
\author{J. K. Freericks}
\affiliation{Department of Physics, Georgetown University, Washington, DC 20057, USA.\\}
\author{A. M. Shvaika}
\affiliation{Institute for Condensed Matter Physics of the National Academy of Sciences of Ukraine, 1 Svientsitskii Street, 79011 Lviv, Ukraine\\}
\enlargethispage*{100pt}
\begin{abstract}
We calculate the resonant inelastic X-ray scattering (RIXS) response in a Mott insulator which is described by the Falicov-Kimball model. The model can be 
solved exactly within the single site dynamical mean-field theory (DMFT) approximation and the calculated RIXS response is accurate up to a local 
background correction. We find that on resonance the RIXS response is greatly enhanced over various other non-resonant background effects and the response 
systematically evolves from a single peak structure, arising due to relaxation processes within the lower Hubbard band, to a two peak structure, arising
due to relaxation processes within the upper Hubbard band as well as across the Mott gap into the lower Hubbard band, as we vary the incident photon 
frequency to access states from the bottom of the lower Hubbard band to the top of the upper Hubbard band. The charge transfer excitations are found to 
disperse monotonically outwards (as a function of transfered energy) as we go from the center of the Brillouin zone towards the zone corner. These 
correlation induced features have been observed by Hasan {\it et. al.} (Science {\bf 288}, 1811 (2000)) and many other experimentalists in RIXS 
measurements over various transition metal oxide compounds and are found to be robust and survive even for large Auger lifetime broadening effects. As a 
comparison, we also calculate the dynamic structure factor for this model, which  is proportional to the nonresonant part of the response, and does not 
show these specific signatures.
\end{abstract}
\pacs{71.10.Fd, 71.27.+a, 74.72.-h, 78.70.Ck}
\maketitle
\section{Introduction}
Resonant inelastic X-ray scattering is essentially a deep core level spectroscopic method which is increasingly becoming an essential technique 
in understanding the complex electronic dynamics of a wide class of novel materials like cuprates, manganites and various other transition metal oxide 
compounds. In the RIXS process, a highly energetic X-ray photon (with energy $\sim 1-10\mbox{ keV}$) excites a deep core level electron into the unoccupied 
states of the conduction band. The excited electron then undergoes inelastic scattering processes with various intrinsic excitations present in the system 
and finally, a conduction band electron fills up the core-hole and emits a photon with relatively lower energy. So, this is a two photon inelastic process 
with no core-hole present in the final state. The transfered energy and momentum to the intrinsic excitations of the system as well as the change in the 
polarization of the scattered photon can provide important information regarding these excitations. Also, RIXS being a resonant technique, the incident 
photon energy can be chosen to coincide with, and hence resonate with, certain intrinsic X-ray atomic transitions which in effect can greatly enhance the 
inelastic scattering cross section and hence can be used as a probe for charge, magnetic and orbital degrees of freedom on selective atomic sites.

RIXS has several advantages over other spectroscopic techniques like angle resolved photo-emission spectroscopy (ARPES) and neutron scattering. First, in 
ARPES, the incident photon knocks out an electron from the system and hence can only probe the occupied states in a system but RIXS, being a high-energy 
process can excite a system into unoccupied intermediate states (like the upper Hubbard band in a Mott insulator) and hence can be used as a probe for 
understanding complex electron dynamics in those strongly correlated intermediate states. Inverse photo-emission techniques, in which an electron is 
injected into the system, can also access the unoccupied states of a system, but this method will charge the system and so far no momentum resolved 
inverse ARPES with sufficient energy resolution has been developed. Secondly, the scattering phase space, \textit{i.e.} the range of energies and momenta 
that can be transfered in the RIXS process, is much larger than other available photon scattering techniques involving visible or infrared light. As a 
result, RIXS can probe low-energy excitations over a wider range of the Brillouin zone and most importantly it can be used to probe all three directions 
of the Brillouin zone, and hence can be used even for materials which are intrinsically three dimensional in nature as compared to the ARPES which, 
because of the in plane momentum conservation, is widely used for materials which are inherently two dimensional in nature and for three dimensional 
materials analysis of ARPES spectra are much more complicated (the perpendicular component of momentum is integrated over in the ARPES spectra). Also, 
photons can much more strongly couple with the electronic system through Coulomb interaction as compared to the neutron-electron interaction arising 
through the spin exchange interaction. As a result, RIXS can be used on very small volume samples, thin films, surfaces and nano-objects in addition to 
bulk single crystal or powder samples. Also, for materials for which growing large volume single crystals is a difficult task, RIXS has significant 
advantage over neutron scattering. Besides these, RIXS is polarization dependent and hence can be used to probe magnetic excitations and also it can be 
used as a probe for certain specific elements or orbitals in a system. The main disadvantage of RIXS over other techniques is, it requires substantially 
large incident photon flux in order to have comparable or better energy and momentum resolution. But recent progress in RIXS instrumentation has 
dramatically improved upon this situation and RIXS is beginning to become an important probe for condensed matter physics.

Over the past decade or so, RIXS has been carried out over large classes of transition metal oxide compounds like cuprates, manganites, irridates 
{\it etc.} (see Ref.~\onlinecite{DevereauxRMP2011} for a detailed review). Most notably, RIXS measurements have been performed at the $\mbox{Cu}$ 
K-edge~\cite{JohnHillPRL1998,AbbamontePRL1999,HasanScience2000,HamalainenPRB2000,HasanPRL2002,KimPRL2002,KimPRL2004,IshiiPRL2005a,IshiiPRL2005b,
JohnHillPRL2008} over a large class of cuprates like the undoped $\mbox{La}_{2}\mbox{CuO}_{4}$,~\cite{AbbamontePRL1999,KimPRL2002} 
$\mbox{Nd}_{2}\mbox{CuO}_{4}$,~\cite{JohnHillPRL1998,HamalainenPRB2000} $\mbox{Sr}_{2}\mbox{CuO}_{2}\mbox{Cl}_{2}$,~\cite{JohnHillPRL1998} 
$\mbox{Ca}_{2}\mbox{CuO}_{2}\mbox{Cl}_{2}$,~\cite{HasanScience2000} quasi one dimensional cuprates like $\mbox{SrCuO}_{2}$,~\cite{HasanPRL2002,KimPRL2004} 
$\mbox{Sr}_{2}\mbox{CuO}_{3}$,~\cite{HasanPRL2002} hole-doped cuprates $\mbox{La}_{2-x}\mbox{Sr}_{x}\mbox{CuO}_{4}$,~\cite{JohnHillPRL2008} 
$\mbox{Y}\mbox{Ba}_{2}\mbox{Cu}_{3}\mbox{O}_{7-\delta}$,~\cite{IshiiPRL2005a} and electron doped cuprates 
$\mbox{Nd}_{1.85}\mbox{Ce}_{0.15}\mbox{CuO}_{4}$,~\cite{IshiiPRL2005b} as well as at the $\mbox{Cu}$ 
$L_{3}$-edge~\cite{GhiringhelliPRL2004,GhiringhelliPRB2007,BraicovichPRL2010} over various undoped cuprates $\mbox{CuO}$,~\cite{GhiringhelliPRL2004} 
$\mbox{Sr}_{2}\mbox{CuO}_{2}\mbox{Cl}_{2}$,~\cite{GhiringhelliPRL2004} $\mbox{La}_{2}\mbox{CuO}_{4}$,~\cite{GhiringhelliPRL2004,BraicovichPRL2010} and 
doped systems $\mbox{Bi}_{2}\mbox{Sr}_{2}\mbox{CaCu}_{2}\mbox{O}_{8+\delta}$, $\mbox{Nd}_{2-x}\mbox{Ce}_{x}\mbox{CuO}_{4}$,~\cite{GhiringhelliPRL2004} 
$\mbox{La}_{2-x}\mbox{Sr}_{x}\mbox{CuO}_{4}$,~\cite{GhiringhelliPRL2004,GhiringhelliPRB2007,BraicovichPRL2010}. Besides these, RIXS measurements at the 
$\mbox{Mn}$ $K$-edge in the orbitally ordered manganite $\mbox{LaMnO}_{3}$,~\cite{InamiPRB2003} at the $\mbox{Mn}$ $\mbox{L}_{2,3}$ absorption edge in 
$\mbox{MnO}$~\cite{GhiringhelliPRB2006} at the $\mbox{Ni}$ $\mbox{L}_{3}$-edge in $\mbox{NiO}$~\cite{GhiringhelliPRL2009} and at the $\mbox{Ir}$ 
$\mbox{L}_{3}$-edge in the $5d$ Mott insulator $\mbox{Sr}_{2}\mbox{IrO}_{4}$~\cite{KimScience2009,IshiiPRB2011} have also been performed. A common feature 
of these materials is that all of them are either Mott insulators or doped Mott insulators and have interesting magnetic ground states. RIXS measurements 
on these materials have probed energy and momentum resolved features of charge transfer excitations~\cite{HasanScience2000,HasanPRL2002,IshiiPRB2011}, 
$dd$ excitations~\cite{GhiringhelliPRL2004,GhiringhelliPRB2007,GhiringhelliPRB2006} (arising due to transitions between crystal field split $d$ orbitals), 
orbitons~\cite{InamiPRB2003} in orbitally ordered systems and even magnetic excitations like magnons~\cite{BraicovichPRL2010} and 
bi-magnons~\cite{JohnHillPRL2008}. 

Theoretical approaches in understanding the RIXS response are mainly either based on exact diagonalization of model Hamiltonians over finite but small 
clusters~\cite{TsutsuiPRL1999,HasanScience2000,VernayPRB2007,VernayPRB2008} or based on a single particle approach~\cite{NomuraJPSJ2004,NomuraJPCS2006,
IgarashiPRB2006,TakahashiJPSJ2008,MarkiewiczPRL2006} which includes realistic band structure effects but treats the correlation effects perturbatively 
under the random phase approximation~\cite{NomuraJPSJ2004,NomuraPRB2005,NomuraJPCS2006,IgarashiPRB2006,TakahashiJPSJ2008} (RPA) or under a self consistent 
renormalization~\cite{MarkiewiczPRL2006} (SCR) approach and the effect of scattering from the core-hole in the first order Born approximation or multiple 
scattering approximation. The exact diagonalization method treats the strong correlation effects exactly but because of the exponentially growing basis 
problem this method is limited to small size clusters and small number of orbitals and hence has limited momentum resolution. Also, the effects of the 
core-hole in this approach as well as in the SCR based calculations are taken either through an input core-hole lifetime, arising due to Auger and 
fluorescence effects, which broadens the intermediate states or under the ultrashort core-hole lifetime (UCL) approximation~\cite{vandenBrinkJPCS2005,
vandenBrinkEPL2006} which is found to be perturbatively exact for small as well as large core-hole potentials. But the effect of the core-hole lifetime 
arising solely due to intrinsic strong correlation effects in a Mott insulator on the RIXS response has not been addressed so far and in this work we use 
the Falicov-Kimball~\cite{FKmodel} (FK) model to address this issue. The main motivation in choosing the FK model is that the FK model is one of the 
simplest models of strongly correlated electron systems which can be exactly solved~\cite{BrandtFK,FreericksRMP2003} under the single site dynamical mean 
field theory~\cite{MetznerPRL1989,KotliarRMP1996} (DMFT) approximation and most notably shows a Mott insulating ground state for large interaction strength 
between the itinerant and the static electrons. Also, the fully renormalized two particle dynamic charge correlation function involving the itinerant 
species as well as the finite temperature core-hole propagator in this model can be calculated exactly. In the following sections, we show that the 
calculated RIXS response in the limit of large core-hole energy is ``exact" provided we : (1) neglect some momentum independent background contributions 
and (2) calculate the charge vertex for exchange processes (see below) under the  Hartree-Fock approximation, which is correct in leading order. We believe 
this model specific calculation can shed some light onto our understanding of the strong correlation effects in the RIXS response and the charge dynamics 
in a Mott insulator in general. The organization of the paper is as follows. In Sec. II, we provide a brief mathematical formulation for the calculation of 
the RIXS cross section followed by Sec. III where we show a more detailed calculation for the RIXS response in the FK model. In Sec. IV, we show our results 
for a half filled Mott insulator followed by Sec. V where we discuss the core-hole lifetime broadening effects on the RIXS response and in Sec. VI we show 
some results for the case of particle-hole asymmetric Mott insulator. Finally, in Sec. VII we conclude.
\section{Mathematical formulation of RIXS}
Our starting point is the familiar electron-photon interaction Hamiltonian~\cite{DevereauxRMP2011} 
\begin{eqnarray}
&&\hspace{-0.2cm} H_{int} = \sum_{i=1}^{N} \left[\frac{e}{m} \mathbf{A}(\mathbf{r}_{i},t)\cdot\mathbf{p}_{i}
                          + \frac{e\hbar}{2m}\boldsymbol{\sigma}_{i}\cdot\boldsymbol{\nabla}\times\mathbf{A}(\mathbf{r}_{i},t)\right . \nonumber \\
&&\hspace{-0.2cm}         + \left . \frac{e^{2}}{2m}\mathbf{A}^{2}(\mathbf{r}_{i},t)   
                          -\frac{e^{2}\hbar}{4m^{2}c^{2}}\boldsymbol{\sigma}_{i}\cdot \frac{\partial \mathbf{A}(\mathbf{r}_{i},t)}{\partial t}\times 
                           \mathbf{A}(\mathbf{r}_{i},t)\right] ,
\label{eq:Hint} 
\end{eqnarray}
for a system of $N$ electrons. $\mathbf{A}(\mathbf{r},t)$ is the vector potential for the external electromagnetic field and can be expanded in a plane 
wave basis as
\begin{eqnarray}
\mathbf{A}(\mathbf{r},t) = \sum_{\mathbf{k},\varepsilon} \sqrt{A_{0}}
\left(\boldsymbol{\varepsilon} a_{\mathbf{k}\varepsilon}e^{i(\mathbf{k}\cdot\mathbf{r}-\omega t)}+
\boldsymbol{\varepsilon}^{*}a_{-\mathbf{k}\varepsilon}^{\dagger}e^{-i(\mathbf{k}\cdot\mathbf{r}-\omega t)}\right),
\end{eqnarray} 
where $A_{0} = \frac{\hbar}{2\mathcal{V}\epsilon_{0}\omega_{\mathbf{k}}}$, $\mathcal{V}$ is the volume of the system, $\boldsymbol{\varepsilon}$ is the 
polarization of the light and we have fixed the gauge by choosing $\boldsymbol{\nabla}\cdot\mathbf{A}(\mathbf{r},t) = 0$ in Eq.~(\ref{eq:Hint}).

In the RIXS process, an incident X-ray photon with momentum $\mathbf{k}_{i}$, energy $\omega_{i}$ and polarization $\boldsymbol{\varepsilon}_{i}$ is 
scattered to a final state described by momentum $\mathbf{k}_{f}$, energy $\omega_{f}$ and polarization $\boldsymbol{\varepsilon}_{f}$. Fermi's golden rule 
to second order in $H_{int}$ gives the transition rate for this process:
\begin{eqnarray}
\mathcal{W} &=& \frac{2\pi}{\hbar}\sum_{\mathbf{F}}\left|\langle \mathbf{F} | H_{int} | \mathbf{I} \rangle +\sum_{n}\frac{\langle \mathbf{F} |H_{int}| n 
\rangle \langle n | H_{int} | \mathbf{I} \rangle}{E_{\mathbf{I}} - \epsilon_{n}}\right|^{2} \nonumber \\ 
& & \hspace*{4cm}\times\delta\left(E_{\mathbf{F}}-E_{\mathbf{I}}\right),
\label{eq:defnW}
\end{eqnarray}
where $| \mathbf{I} \rangle \equiv |i \rangle \otimes |\mathbf{k}_{i},\omega_{i}\rangle$, $|\mathbf{F} \rangle \equiv |f \rangle \otimes |\mathbf{k}_{f},
\omega_{f}\rangle$ and $|n \rangle$ are the matrix product states for the initial, final and the intermediate states of the systems respectively (both 
electronic and photon states are present in the initial and final states while the intermediate states are just the electronic states) and $E_{\mathbf{I}}$, 
$E_{\mathbf{F}}$ and $\epsilon_{n}$ are the corresponding energies, respectively. It is interesting to mention that the intermediate state $|n\rangle$ has 
a core-hole while the initial and final states, $|\mathbf{I}\rangle$ and $|\mathbf{F}\rangle$ have no core-hole. The first-order amplitude is in general 
dominant over the second-order contribution except near resonance when the incident photon energy is nearly equal to a specific atomic transition in a 
material; \textit{i. e.}, $\omega_{i} \approx \epsilon_{n} - \epsilon_{i}$. At resonance, the second-order term becomes overwhelmingly large compared to the 
first-order term and hence the second-order term causes resonant scattering while the first order term gives rise to nonresonant scattering.

The diamagnetic term proportional to $\mathbf{A}^{2}$ as well as the spin-orbit coupling term proportional to 
$\boldsymbol{\sigma}\cdot (\partial \mathbf{A}/\partial t)\times \mathbf{A}$ in Eq.~(\ref{eq:Hint}) contribute to the first-order amplitude. The latter is 
smaller than the former by a factor of $\omega_{i(f)}/mc^{2}\ll 1$ and also, at resonance, the contribution from the diamagnetic term is negligibly small 
compared to the resonant term and hence their contribution will be neglected. So, then the resonant part of the second-order amplitude at zero temperature 
is given by~\cite{DevereauxRMP2011,PlatzmanPRB1998,NozieresPRB1974}   
\begin{eqnarray}
&&\frac{e^{2}\hbar\sqrt{\omega_{\mathbf{k}_{i}}\omega_{\mathbf{k}_{f}}}}{\mathcal{V}\epsilon_{0}}
\sum_{n}\left[\frac{\langle f |\mathcal{D}_{\mathbf{k}_{f}}|n\rangle \langle n |\mathcal{D}^{\dagger}_{-\mathbf{k}_{i}}|i\rangle}
{\varepsilon_{n}-\varepsilon_{i}-\omega_{i}}\right . \nonumber \\
&&\hspace*{1cm}\left. +\frac{\langle f |\mathcal{D}^{\dagger}_{-\mathbf{k}_{i}}|n\rangle \langle n |\mathcal{D}_{\mathbf{k}_{f}}|i\rangle}
{\varepsilon_{n}-\varepsilon_{i}+\omega_{f}} \right]\delta\left(\epsilon_{f}-\epsilon_{i}-\Omega\right)
\label{eq:2ndorderAmplitude}
\end{eqnarray}
where $\Omega = \omega_{i}-\omega_{f}$ is the transfered energy and 
\begin{eqnarray}
\mathcal{D}_{\mathbf{k}}=\frac{1}{im\omega_{\mathbf{k}}}\sum_{i=1}^{N} e^{i\mathbf{k}\cdot\mathbf{r}_{i}}
\left(\boldsymbol{\varepsilon}\cdot \mathbf{p}_{i}+\frac{i\hbar}{2}\boldsymbol{\sigma}_{i}\cdot\mathbf{k}\times\boldsymbol{\varepsilon}\right),
\label{eq:defnofD}
\end{eqnarray}
is the relevant transition operator for the RIXS cross section. The first term in Eq.~(\ref{eq:defnofD}) causes nonmagnetic scattering while the second 
term, arising from the spin-orbit coupling term in $H_{int}$, causes magnetic scattering which, for typical incident photon energy ($\sim 1-10 \mbox{keV}$)
and the localized core levels involved in a RIXS process, is about 100 times smaller than the non-magnetic term~\cite{DevereauxRMP2011} and hence will also 
be neglected. Finally, under such circumstances, we assume the dipole limit for the RIXS process and the transition operator is then given by
\begin{eqnarray}
\mathcal{D}=\boldsymbol{\varepsilon}\cdot\mathbf{D} \hspace{0.25cm} \mbox{with} \hspace{0.25cm} \mathbf{D}=\frac{1}{im\omega_{\mathbf{k}}}
\sum_{i=1}^{N} \mathbf{p}_{i}
\end{eqnarray}
\section{RIXS response in the Falicov-Kimball model}
The single-site Hamiltonian of the Falicov-Kimball model (in the hole representation) including the interaction with a core-hole is given by 
\begin{eqnarray}
H_{loc} &=& Un_{d}n_{f}+Q_{d}n_{d}n_{h}+Q_{f}n_{f}n_{h}-\mu n_{d}\nonumber \\
& &\hspace*{1.2cm}+(E_{f}-\mu)n_{f}+(E_{h}-\mu)n_{h}
\end{eqnarray}
where $n_{d}=d^{\dagger}d$, $n_{f}=f^{\dagger}f$ and $n_{h}=h^{\dagger}h$ are the occupation number operators for the $d$-hole, $f$-hole and core-hole 
state, respectively. $U$ is the onsite Coulomb interaction between the itinerant $d$ and the static $f$ holes, $Q_{d} > 0$ and $Q_{f} > 0$ are the 
Coulomb interactions between the core-hole and the $d$-hole and $f$-hole, respectively, $E_{f}$ is the site energy of the $f$ state and 
$E_{h}\sim 10^{2}-10^{4}$ eV is the energy of the core-hole state.

\begin{figure}
\includegraphics[scale=1.0,clip]{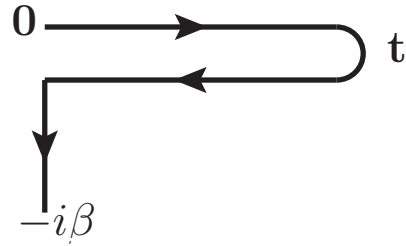}
\caption{The Kadanoff-Baym-Keldysh contour. The contour starts at time 0, moves forward along the real axis to time $t$ then moves backward along the 
real axis to time 0 and finally downwards along the imaginary axis to time $-i\beta$.}
\label{fig:KeldyshContour}
\end{figure}

The full Hamiltonian on the lattice includes a repeat of this local Hamiltonian for each lattice site and a hopping of the itinerant $d$-holes between 
neighboring sites. The density matrix for the single impurity problem in DMFT is then given by 
\begin{eqnarray}
\rho = \frac{e^{-\beta H_{loc}}}{\mathcal{Z}} \mathcal{T}_{c}\exp\left\{ - i\!\int_{c} dt' \int_{c} \! dt'' d^{\dagger}(t')\lambda(t',t'')d(t'')\right\},
\label{eq:defnrho}
\end{eqnarray}  
where the time-ordering and integration are performed over the Kadanoff-Baym-Keldysh contour as shown in Fig.~\ref{fig:KeldyshContour} and 
$\beta=1/k_{B} T$ is the inverse temperature.
Here, the dynamical mean-field $\lambda_{c}(t',t'')$ and the chemical potential $\mu$ are taken from the equilibrium solution of the conduction electron 
problem without the core-hole, arising under the single site DMFT approximation. This in effect implies that we are treating the creation of the core-hole 
under the sudden approximation instead of a full non-equilibrium treatment of the core-hole propagator. 

The \textit{impurity problem} arising under the DMFT approximation can be solved \textit{exactly} in this case and the \textit{local} $d$-hole propagator, 
$G_{d}^{loc}(\omega)$, is given by
\begin{eqnarray}
G_{d}^{loc}(\omega) = \frac{w_{0}}{\omega^{+}+\mu-\lambda(\omega^{+})}+\frac{w_{1}}{\omega^{+}+\mu-\lambda(\omega^{+})-U},
\end{eqnarray}
where $\omega^{\pm}=\omega \pm i\delta$ ($\delta > 0$), $w_{0}$ and $w_{1}$ are the probabilities for finding a given site unoccupied and occupied by an 
$f$-hole,respectively. The momentum dependent fully renormalized $d$-hole propagator is given by 
\begin{eqnarray}
G_{d}(\mathbf{q},\omega)=\frac{1}{\omega^{+}+\mu-\epsilon_{\mathbf{q}}-\Sigma_{d}^{loc}(\omega^{+})},
\end{eqnarray}
where the \textit{local} self energy is related to the \textit{local} propagator through Dyson's equation
\begin{eqnarray}
\Sigma_{d}^{loc}(\omega^{+})=\omega^{+}+\mu-\lambda(\omega^{+})-\left[G_{d}^{loc}(\omega^{+})\right]^{-1}.
\end{eqnarray}

Similarly, the core-hole Green's functions, $G_{h}^{>}(t)=-i\langle h(t)h^{\dagger}(0)\rangle$ and $G_{h}^{<}(t)=i\langle h(t)h^{\dagger}(0)\rangle$ can 
also be calculated~\cite{NP-JKF-AScorehole} by using either numerical integration over the Kadanoff-Byam-Keldysh contour or by the Wiener-Hopf sum 
equation~\cite{WHapproach,ShvaikaWH} approach [the angular brackets $\langle \; \rangle$ denote a trace over all states weighted by density matrix in 
Eq.~\ref{eq:defnrho} and the operators are in the interaction representation with respect to $H_{loc}$]. Also, it is important to mention that, for the 
calculation of the itinerant as well as the core-hole propagators we use the $d$ dimensional hypercubic lattice density of states (DOS) in the limit of 
$d\rightarrow\infty$ (DMFT approximation).

\begin{figure}
\includegraphics[scale=0.6,clip]{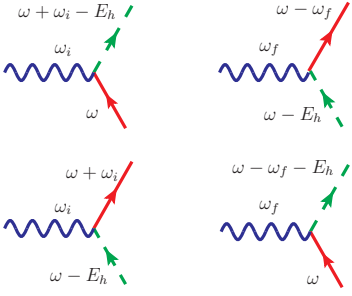}
\caption{(Color online) RIXS interaction vertices. Wavy lines (blue) represent incident or scattered photons, the dashed lines (green) represent the 
core-hole propagator, and the solid lines (red) represent the propagator for the itinerant $d$-holes. The labels indicate the energies of the different 
particles. Note that, the momentum and energy are conserved at each vertex.}
\label{fig:RIXSinteraction}
\end{figure}

The interaction of the X-ray photon with the electronic subsystem of matter can be represented by the diagrams shown in Fig.~\ref{fig:RIXSinteraction}.
We have explicitly shown the direct dependence of the core-hole propagator on the core-hole energy $E_{h}$, which is supposed to be much larger than the 
band energies and is of the order of the incident ($\omega_{i}$) and the scattered X-ray photon energies ($\omega_{f}$), respectively. One can see that in 
the case of large photon and core-hole energies only the first two vertices (top two diagrams in Fig.~\ref{fig:RIXSinteraction}) contribute significantly 
whereas the contribution from the remaining two (bottom two diagrams in Fig.~\ref{fig:RIXSinteraction}) are negligibly small because the hole-propagators 
are evaluated too far off the energy shell.

\begin{figure}
\includegraphics[scale=0.6,clip]{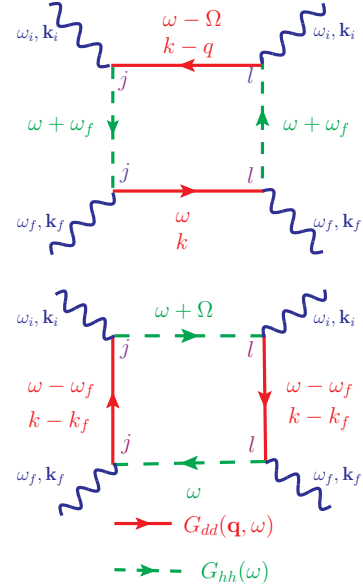}
\caption{(Color online) Bare loop contribution to the direct RIXS process. The bottom diagram gives a momentum-independent contribution (see text) to the 
RIXS process and hence will be neglected in this calculation.}
\label{fig:DirectRIXSminimal}
\end{figure}

In that limit, the bare-loop contribution to the amplitude for the RIXS process is represented by the two diagrams shown in 
Fig.~\ref{fig:DirectRIXSminimal}.  
The contribution of the top diagram to the four-particle correlation function, $\chi^{(4)}_{i,f,f,i}(i\nu_{i},i\nu_{f},i\nu'_{f},i\nu'_{i})$, evaluated on 
the imaginary axis, is equal to
\begin{eqnarray}
-\frac{1}{\beta}\sum_{m}\chi_{0}^{dd}(i\omega_{m}-i\nu_{i}+i\nu_{f},i\omega_{m}|\mathbf{q})\nonumber \\ 
\times\chi_{0}^{hh}(i\omega_{m}+i\nu_{f},i\omega_{m}+i\nu'_{f})
\end{eqnarray}
whereas the contribution of the bottom diagram to $\chi^{(4)}_{i,f,f,i}(i\nu_{i},i\nu_{f},i\nu'_{f},i\nu'_{i})$ is equal to
\begin{eqnarray}
-\frac{1}{\beta}\sum_{m}\chi_{0}^{hh}(i\omega_{m}+i\nu_{i}-i\nu_{f},i\omega_{m})\nonumber \\ 
\times\chi_{0}^{dd}(i\omega_{m}-i\nu_{f},i\omega_{m}-i\nu'_{f}|0),
\label{eq:LocalContrbn}
\end{eqnarray}
where we have introduced the bare charge susceptibilities 
\begin{eqnarray}
&&\chi_{0}^{dd}(i\omega_{m}+i\nu,i\omega_{m}|\mathbf{q}) = -\frac{1}{N}\sum_{\mathbf{k}}G_{d}(\mathbf{k}+\mathbf{q},i\omega_{m}+i\nu)\nonumber \\
&&\hspace*{5cm} \times\;G_{d}(\mathbf{k},i\omega_{m}),\\
&&\chi_{0}^{hh}(i\omega_{m}^{h}+i\nu,i\omega_{m}^{h}) = -G_{h}(i\omega_{m}^{h}+i\nu)G_{h}(i\omega_{m}^{h})
\end{eqnarray}
and $i\omega_{m}^{h}\equiv i\omega_{m}-E_{h}$. Since the core-hole propagator is local, the bottom diagram in Fig.~\ref{fig:DirectRIXSminimal} does not 
depend on the photon wavevector and hence can only contribute to momentum-independent background effects as evident in Eq.~(\ref{eq:LocalContrbn}), 
whereas the first diagram depends on the transfered momentum $\mathbf{q}=\mathbf{k}_{i}-\mathbf{k}_{f}$. As in the present study we are interested in the 
energy and wavevector dependence of the RIXS response, we neglect all such momentum-independent contributions. The technical reason behind neglecting these 
terms is that the momentum independent contributions are the local ones and hence they include all types of many body scattering processes and can only be 
derived from the solution of the single-impurity problem for the four-particle correlation function $\chi^{(4)}$ (which involves multiparticle vertices and 
many more complications) and at this moment we do not have well developed approach for this. 

It is important to mention that RIXS processes~\cite{vandenBrinkJPCS2005,vandenBrinkEPL2006} can happen either through a \textit{direct} process, as in the 
case of the $L_{2,3}$-edge $2p \rightarrow 3d$ RIXS, in which the core-electron is excited to an unoccupied state of the correlated valence band ($d$-band) 
or through an \textit{indirect} process, as in the case of $K$-edge $1s \rightarrow 4p$ RIXS, in which the excited core-electron goes into an uncorrelated 
$4p$ band several eV above the Fermi level. In the present case, the excited core-electron goes to the correlated $d$-band and dipole selective transition
 $2p \rightarrow 3d$ or $3p \rightarrow 4d$ is consistent with the involvement of a $2p$ or $3p$ core-hole and hence our study is related to the direct 
RIXS processes like $L$-edge or $M$-edge RIXS. In particular, we will be studying the $L$-edge RIXS in the following sections.

The calculation of the RIXS response involves analytic continuation to real frequencies, which is a well defined but tedious 
procedure~\cite{ShvaikaPRB2005} and gives the following contribution to the RIXS cross section,  
\begin{eqnarray}
&&\hspace*{-0.35cm}-\frac{1}{2\pi^{2}}\int_{-\infty}^{+\infty} d\omega \left[f(\omega)-f(\omega+\Omega)\right]
\chi_{0}^{hh}(\omega^{-}+\omega_{i},\omega^{+}+\omega_{i})\nonumber \\ 
&&\hspace*{0.1cm}\times\mbox{Re}\left[\chi_{0}^{dd}(\omega^{-},\omega^{-}+\Omega|\mathbf{q})
-\chi_{0}^{dd}(\omega^{+},\omega^{-}+\Omega|\mathbf{q})\right],
\label{eq:DirectRIXSbare}
\end{eqnarray}
where $f(\omega)=1/[\exp(\beta\omega)+1]$ is the Fermi function and
\begin{eqnarray}
\chi_{0}^{hh}(\omega^{-}+\omega_{i},\omega^{+}+\omega_{i}) = -\left|G_{h}(\omega^{+}+\omega_{i}-E_{h})\right|^{2}.
\end{eqnarray}

Similary, the contribution from the second diagram is given by
\begin{eqnarray}
& &\frac{1}{2\pi^{2}}\int_{-\infty}^{+\infty}\left[f(\omega)-f(\omega+\Omega)\right]  
\chi_{0}^{dd}(\omega^{-}+\omega_{f},\omega^{+}+\omega_{i}|0)\nonumber \\ 
& &\hspace*{0.5cm}\times\mbox{Re}\left[\chi_{0}^{hh}(\omega^{-}-\Omega,\omega^{-})
-\chi_{0}^{hh}(\omega^{+}-\Omega,\omega^{-})\right]
\end{eqnarray}
and as already has been mentioned it does not depend on the photon momentum and hence will be neglected. 

\begin{figure}
\begin{center}
\includegraphics[scale=0.5,clip]{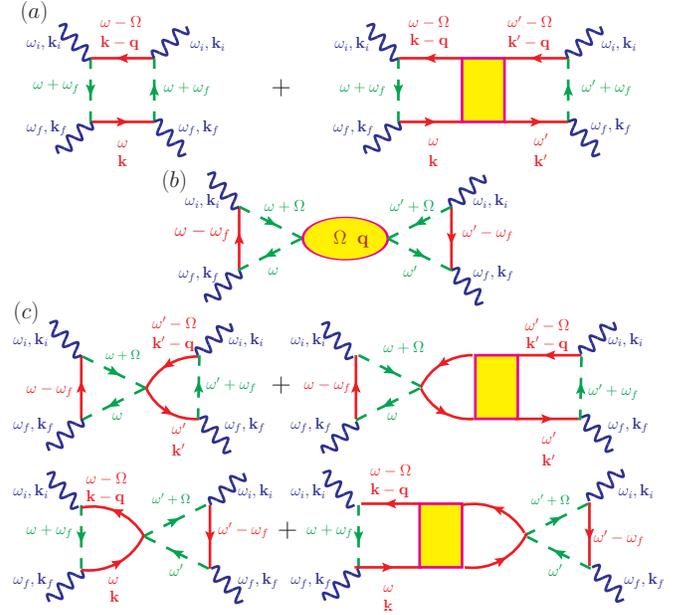}
\caption{(Color online) Diagrams for the RIXS response. Panel (a) total \textit{direct resonant scattering}, (b) \textit{full exchange resonant scattering} 
and (c) \textit{partial exchange resonant scattering} processes. Explicit $E_{h}$ dependence of the core-hole propagator is not shown in the figure.}
\label{fig:TotalRIXSresponse}
\end{center}
\end{figure}

Next, we introduce the renormalization of the bare charge susceptibilities through inclusion of charge vertices. In the simplest case, this can be done by 
inserting the two-particle charge vertex either in between the two $d$-hole propagators ($dd$-channel) or between the two core-hole propagators ($hh$-channel). 
First, we consider the effect of charge screening in the $dd$-channel which is shown in the second diagram in Fig.~\ref{fig:TotalRIXSresponse} (a). 
The sum of the two diagrams in Fig.~\ref{fig:TotalRIXSresponse}(a) corresponds to the replacement in Eq.~(\ref{eq:DirectRIXSbare}) of the bare charge 
susceptibility $\chi_{0}^{dd}(i\omega_{m}+i\nu,i\omega_{m}|\mathbf{q})$ by the fully renormalized charge susceptibility 
$\chi^{dd}(i\omega_{m}+i\nu,i\omega_{m}|\mathbf{q})$ which in the case of Falicov-Kimball model is given by 
\begin{eqnarray}
& &\left[\chi^{dd}(i\omega_{m}+i\nu,i\omega_{m}|\mathbf{q})\right]^{-1}=\left[\chi_{0}^{dd}(i\omega_{m}+i\nu,i\omega_{m}|\mathbf{q})\right]^{-1}\nonumber \\
& & \hspace*{4.5cm}+\Gamma(i\omega_{m}+i\nu,i\omega_{m}),
\end{eqnarray}
where the irreducible charge vertex~\cite{ShvaikaPhysicaC2000,FreericksPRB2000} is
\begin{eqnarray}
\Gamma(i\omega_{m}+i\nu,i\omega_{m})=\frac{1}{T}\frac{\Sigma(i\omega_{m})-\Sigma(i\omega_{m}+i\nu)}{G(i\omega_{m})-G(i\omega_{m}+i\nu)}.
\end{eqnarray}
The sum of the two diagrams in Fig.~\ref{fig:TotalRIXSresponse}(a) corresponds to the \textit{direct scattering} contribution to the RIXS process and is 
given by
\begin{eqnarray}
&&\hspace*{-0.25cm}-\frac{1}{2\pi^{2}}\int_{-\infty}^{+\infty} d\omega \left[f(\omega)-f(\omega+\Omega)\right]
\left|G_{h}(\omega^{+}+\omega_{i}-E_{h})\right|^{2}\nonumber \\ 
&&\times\mbox{Re}\left[\chi^{dd}(\omega^{-},\omega^{-}+\Omega|\mathbf{q})
-\chi^{dd}(\omega^{+},\omega^{-}+\Omega|\mathbf{q})\right].
\label{eq:DirectRIXSrenorm}
\end{eqnarray}
The word direct here refers to a direct Coulomb scattering process since we already discussed that we will calculate only direct RIXS process in this work.
The contribution of the diagram in Fig.~\ref{fig:TotalRIXSresponse}(b), which we label as the \textit{full exchange scattering} processes to the 
four-particle correlation function, $\chi^{(4)}_{i,f,f,i}(i\nu_{i},i\nu_{f},i\nu'_{f},i\nu'_{i})$, on the imaginary axis is given by
\begin{eqnarray}
Q_{d}^{2}\Lambda(i\nu_{i},i\nu_{f},i\nu_{i}-i\nu_{f})\Lambda(i\nu'_{i},i\nu'_{f},i\nu_{i}-i\nu_{f})\nonumber \\ 
\times\chi^{dd}(i\nu_{i}-i\nu_{f},\mathbf{q}),
\label{eq:IndirectRIXS}
\end{eqnarray}
where
\begin{eqnarray}
&&\hspace*{-0.25cm}\chi^{dd}(i\nu,\mathbf{q})=\frac{1}{\beta}\sum_{m}\chi^{dd}(i\omega_{m}+i\nu,i\omega_{m}|\mathbf{q})\nonumber\\
&&\hspace*{-0.25cm}=\frac{i}{2\pi}\int_{-\infty}^{+\infty}d\omega
\left[\chi^{dd}(\omega^{+},\omega+i\nu|\mathbf{q})-\chi^{dd}(\omega^{-},\omega+i\nu|\mathbf{q})\right . \nonumber\\
&&\hspace*{-0.25cm}\left .+\chi^{dd}(\omega-i\nu,\omega^{+}|\mathbf{q})-\chi^{dd}(\omega-i\nu,\omega^{-}|\mathbf{q})\right]f(\omega),
\end{eqnarray}
is the dynamical charge susceptibility and $\Lambda$ satisfies
\begin{eqnarray}
\Lambda(i\nu_{i},i\nu_{f},i\nu_{i}-i\nu_{f})&=&-\frac{1}{\beta}\sum_{m}G_{h}(i\omega_{m}+i\nu_{i}-E_{h})\nonumber \\
&&\hspace*{-0.8cm}\times G_{d}(i\omega_{m})G_{h}(i\omega_{m}+i\nu_{f}-E_{h}).
\end{eqnarray}
It is important to mention that in the derivation of the full exchange scattering process contribution to RIXS in Eq.~(\ref{eq:IndirectRIXS}), we 
have approximated the core-hole-$d$-hole charge vertex by the core-hole - $d$-hole interaction $Q_{d}$ under the Hartree-Fock approximation, which is 
exact to leading order. After analytic continuation to real frequencies, the corresponding contribution to the RIXS cross section is equal to
\begin{eqnarray}
\frac{Q_{d}^{2}}{\pi}\mbox{Im}\left[\Lambda(\omega_{i}+i\delta,\omega_{f}+i\delta,\Omega+i\delta)\right . \nonumber \\
\left . \times\Lambda(\omega_{i}-i\delta,\omega_{f}-i\delta,\Omega+i\delta)\chi^{dd}(\Omega+i\delta,\mathbf{q})\right],
\end{eqnarray}
where $\Lambda(\omega_{i}\pm i\delta,\omega_{f}\pm i\delta,\Omega+i\delta)$ is given by
\begin{eqnarray}
&&\hspace*{-0.25cm}\frac{1}{\pi}\int_{-\infty}^{+\infty}d\omega \; f(\omega)\left[G_{h}(\omega^{\pm}+\omega_{i}^{h})G_{h}(\omega^{\pm}+\omega_{f}^{h})
\mbox{Im} G_{d}(\omega^{+})\right . \nonumber \\
&&\hspace*{-0.25cm}\left. +G_{d}(\omega^{\mp}-\omega_{i})G_{h}(\omega^{\mp}-\Omega-E_{h})\mbox{Im}G_{h}(\omega^{+}-E_{h})\right.\nonumber \\
&&\hspace*{-0.25cm}\left. +G_{d}(\omega^{\mp}-\omega_{f})G_{h}(\omega^{\pm}+\Omega-E_{h})\mbox{Im}G_{h}(\omega^{+}-E_{h})\right]
\end{eqnarray}
and $\omega_{i,f}^{h}\equiv\omega_{i,f}-E_{h}$ are the incident and scattered photon energies measured with respect to the core-hole energy, $E_{h}$.

For large core-hole energy, $E_{h}$, we need to keep only the first term which has a small difference of energies $\mu+\omega_{i,f}-E_{h}$ and can safely 
neglect the other two terms containing a large differences in energies ($\mu-E_{h}$ and $\mu+\Omega-E_{h}$, respectively). We then obtain the contribution 
from the full exchange processes as 
\begin{eqnarray}
\frac{Q_{d}^{2}}{\pi}\left|\bar{\Lambda}(\omega_{i}+i\delta,\omega_{f}+i\delta)\right|^{2} \mbox{Im}\chi^{dd}(\Omega+i\delta|\mathbf{q})
\end{eqnarray} 
where,
\begin{eqnarray}
\bar{\Lambda}(\omega_{i}+i\delta,\omega_{f}+i\delta)&=&\frac{1}{\pi}\int_{-\infty}^{+\infty}d\omega f(\omega)
G_{h}(\omega^{+}+\omega_{i}^{h})\nonumber \\
& &\hspace*{-0.5cm}\times G_{h}(\omega^{+}+\omega_{f}^{h})\mbox{Im} G_{d}(\omega^{+}),\\
\mbox{Im}\chi^{dd}(\Omega+i\delta,\mathbf{q}) &=& \frac{1}{2\pi} \int_{-\infty}^{+\infty} d\omega \left[f(\omega)-f(\omega+\Omega)\right]\nonumber \\
& &\hspace*{-3.2cm}\times\mbox{Re}\left[\chi^{dd}(\omega^{-},\omega^{-}+\Omega|\mathbf{q})-\chi^{dd}(\omega^{+},\omega^{-}+\Omega|\mathbf{q})\right].
\end{eqnarray}

Apart from the \textit{direct} and \textit{full exchange} resonant scattering processes, we also have processes, termed as the \textit{partial exchange} 
processes, as shown in Fig.~\ref{fig:TotalRIXSresponse}(c) [this process arises due to the cross terms in the square in Eq.~(\ref{eq:defnW})]. The sum of 
their contributions to the four-particle correlation function, $\chi^{(4)}_{i,f,f,i}(i\nu_{i},i\nu_{f},i\nu'_{f},i\nu'_{i})$, on the imaginary axis, is 
equal to 
\begin{eqnarray}
-Q_{d}\frac{1}{\beta}\sum_{m}\chi^{dd}(i\omega_{m},i\omega_{m}+i\nu_{i}-i\nu_{f}|\mathbf{q})\nonumber \\
\times\left[G_{h}(i\omega_{m}+i\nu'_{i}-E_{h})\Lambda(i\nu_{i},i\nu_{f},i\nu_{i}-i\nu_{f})\right . \nonumber \\
\left . +G_{h}(i\omega_{m}+i\nu_{i}-E_{h})\Lambda(i\nu'_{i},i\nu'_{f},i\nu'_{i}-i\nu'_{f})\right].
\end{eqnarray}
After analytic continuation to real frequencies, we obtain the following partial exchange contribution to the RIXS response
\begin{eqnarray}
& &\hspace*{-0.3cm}-\frac{Q_{d}}{\pi^{2}}\mbox{Re}\left\{\bar{\Lambda}(\omega_{i}+i\delta,\omega_{f}+i\delta)
\int_{-\infty}^{+\infty}d\omega \left[f(\omega)-f(\omega+\Omega)\right]\right . \nonumber \\ 
& &\hspace*{1cm}\left . \times G_{h}(\omega^{-}+\omega_{i}-E_{h})
\mbox{Re}\left[\chi^{dd}(\omega^{-},\omega^{-}+\Omega|\mathbf{q})\right . \right . \nonumber \\ 
& & \hspace*{3.8cm}\left . \left . -\chi^{dd}(\omega^{+},\omega^{-}+\Omega|\mathbf{q})\right]\right\}.
\end{eqnarray} 

\begin{figure}[htbp]
\includegraphics[scale=0.6,clip]{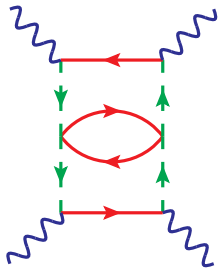}
\caption{(Color online) Parquet diagrams with vanishing contribution to the RIXS response in the limit of infinite dimension.}
\label{fig:vanishingDiagram}
\end{figure}

It is interesting to mention that for the $L$-edge RIXS process, which we consider in the present study, pure direct resonant scattering processes are 
overwhelmingly dominant over the resonant exchange (both full and partial) processes. It is also interesting to mention that in Fig. 4(a) inclusion of the 
reducible charge vertex, which in general is nonlocal, into the core-hole-core-hole channel will give a vanishing contribution to the RIXS cross-section. 
This is because of the local nature (momentum independence) of the core-hole propagator and the vanishing of the $dd$-bubble (charge susceptibility) in 
the diagrams like the one presented in Fig.~\ref{fig:TotalRIXSresponse}(b), but aligned (rotated) in the vertical direction (with the external photon 
lines kept unchanged), in the uniform limit ($\mathbf{q} = 0$). Besides the diagrams considered above, we could also have considered other contributions 
to the charge vertex between the two core-hole propagators which produce diagrams like the parquet diagram in Fig.~\ref{fig:vanishingDiagram}. From a 
simple power counting argument we can show that the contribution of such diagrams goes at least as an inverse power of the dimension of the lattice, $d$. 
So, in the limit of $d \to \infty$, they all have vanishing contributions except in the case when they are all local (and we neglect all such momentum-
independent contributions).

We could also have included renormalization through the core-hole loops
\begin{eqnarray}
\hspace*{-0.25cm}\chi_{0}^{hh}(i\nu) &\equiv& \frac{1}{\beta}\sum_{m}\chi_{0}^{hh}(i\omega+i\nu,i\omega_{m}) \nonumber \\
\hspace*{-0.25cm}&=&\frac{1}{\pi}\int_{-\infty}^{+\infty} d\omega f(\omega) \mbox{Im} G_{h}(\omega^{+}-E_{h})\nonumber \\
&\times&\left[G_{h}(\omega+i\nu-E_{h})+G_{h}(\omega-i\nu-E_{h})\right],
\end{eqnarray}
but they also contain large differences of energies $\mu-E_{h}$ and $\mu+\Omega-E_{h}$ and hence can be neglected for large core-hole energy, $E_{h}$.

The nonresonant part of the RIXS response is found to be related to the density-density correlation function. To be precise, the nonresonant part is 
proportional to the \textit{dynamical structure factor}~\cite{DevereauxPRB2003}, $S(\mathbf{q},\Omega)$ which is given by 
\begin{eqnarray}
S(\mathbf{q},\Omega) = -\frac{1}{\pi}\left[1+n_{B}(\Omega,T)\right]\mbox{Im} \chi^{dd}(\Omega+i\delta|\mathbf{q}),
\end{eqnarray}
where $n_{B}(\omega,T)=1/[\exp(\beta\omega)-1]$ is the Bose distribution function and $\chi^{dd}(\Omega+i\delta|\mathbf{q})$, is the dynamic charge 
susceptibility of the system.
\section{RIXS response for the half filled Mott insulator}
As has been already stated, the Falicov-Kimball model at half filling ($n_{f}=0.5$) shows a Mott insulating ground state for $U > U_{c}=\sqrt{2}$. We 
choose $U=2.0$, $Q_{d}=Q_{f} = 2.5$ and $T = 0.1$ in units of effective hopping amplitude $t^{*}$ and from here onwards we choose $t^{*}=1$. This choice of 
$U$ gives a small ``gap" ($\Delta_{gap} \simeq 0.25$ in units of $t^{*}$) Mott insulator. It is interesting to mention that for the $d\rightarrow\infty$ 
hypercubic lattice DOS there is no true gap as there is an exponentially small DOS inside the gap. 
\begin{figure}
\includegraphics[scale=0.3,clip]{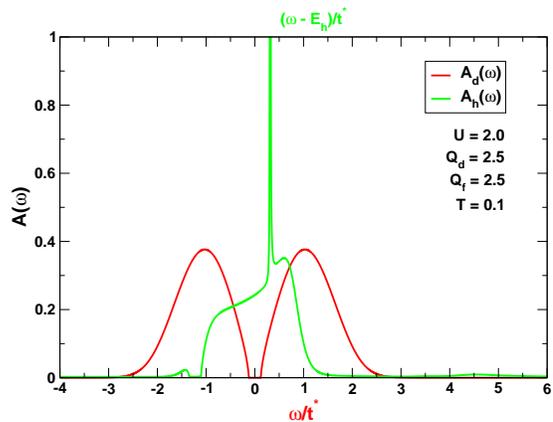}
\caption{(Color online) Spectral function for the $d$-hole, $A_{d}(\omega)$, and the core-hole, $A_{h}(\omega)$. For $A_{h}(\omega)$, the frequency is 
measured with respect to the core-hole energy, $E_{h}$, as shown in the top horizontal axis.}
\label{fig:SpectralFunction}
\end{figure}

In Fig.~\ref{fig:SpectralFunction}, we show the spectral function for the itinerant species, $A_{d}(\omega)$ as well as for the core-hole, 
$A_{h}(\omega-E_{h})$. $A_{d}(\omega)$ clearly shows a ``gap" (Mott gap) at the Fermi level ($\omega = 0$) while $A_{h}(\omega-E_{h})$ also shows a ``gap''
 at some other frequency and the origin of this gap is related to the same strong correlation effects that gives rise to the Mott gap in the itinerant 
species spectral function. Surprisingly, the gap structure is quite different for the core-hole, which is arising due to the asymmetry in the Green's 
function for large $E_{h}$. $A_{h}(\omega)$ is dominated by a broad feature, arising from the projection of $G^{>}(t)$ onto the $n_{h}=0, n_{f}=0$ 
configuration, along with a very sharp peak arising from the projection onto the $n_{h}=0, n_{f}=1$ configuration in the final state.
    
From the knowledge of the itinerant electron propagator, $G_{d}(\mathbf{q},\omega)$, core-hole propagator, $G_{h}(\omega)$ and the fully renormalized 
two-particle charge susceptibility, $\chi^{dd}(i\omega_{m},i\omega_{m}+i\nu|\mathbf{q})$, we can calculate the RIXS response either as a function of 
transfered energy ($\Omega$) for a given fixed incident photon energy, $\omega_{i}$ (measured with respect to the core-hole energy $E_{h}$), or as a 
function of $\omega_{i}$ for a given fixed transfered energy, $\Omega$, for various transfered momenta $\mathbf{q}$ of the photon. It is interesting 
to mention that in the limit $d\rightarrow\infty$ the momentum on the hypercubic lattice only enters through the dimensionless 
parameter~\cite{MullerHartmanZPB1989},
\begin{eqnarray}
X(\mathbf{q})=\lim\limits_{d\rightarrow\infty}\frac{1}{d}\sum_{i=1}^{d}\cos(q_{i}).
\end{eqnarray}
So, $-1\leq X \leq 1$ and $X=1$ and $X=-1$ corresponds to the center, $(0,\cdots,0)$ ($\Gamma$ point), and the corner, $(\pi,\cdots,\pi)$ (M point), of 
the Brillouin zone of a $d$-dimensional hypercubic lattice, respectively. It is convenient to think of this parametrization as corresponding to RIXS 
scattering in the diagonal $\langle 1 \cdots 1 \rangle$ direction.  

\begin{figure}
\includegraphics[scale=0.9,clip]{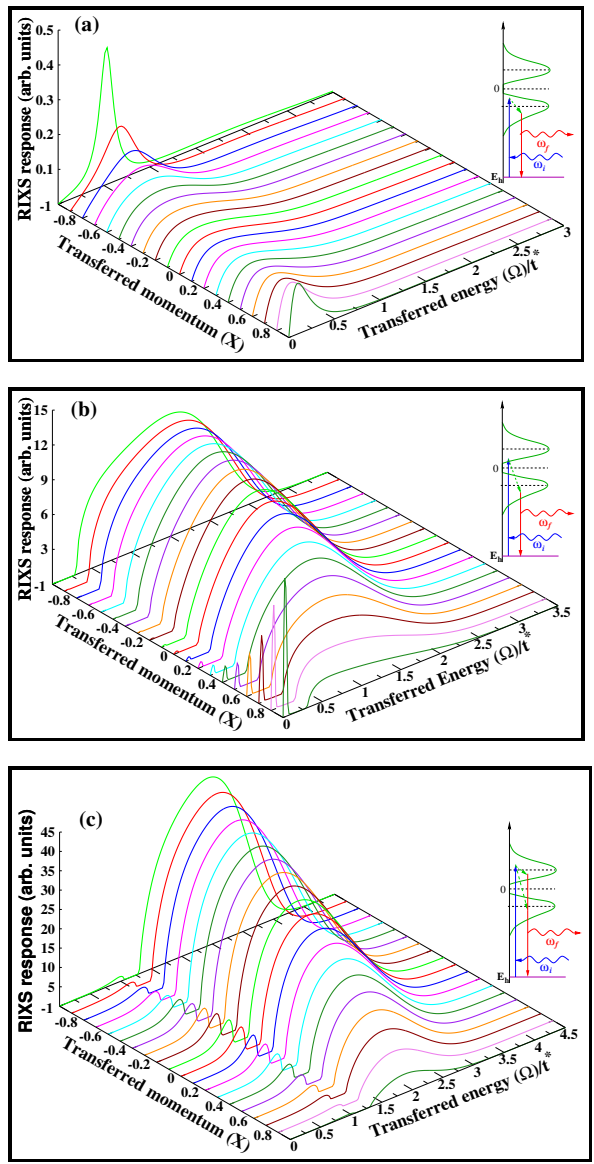}
\caption{(Color online) RIXS response as a function of transfered energy ($\Omega$) for various fixed transfered momenta ($X$) with three different 
incident photon energies, $\omega_{i}$. In panel (a), the incident photon energy is at the lower Hubbard band, in (b) at the bottom of the upper Hubbard 
band and in (c) it is at the middle of the upper Hubbard band. The inset of each plot schematically depicts the possible RIXS relaxation processes. The 
chosen parameters are $U=2.0$, $Q_{d}=Q_{f}=2.5$ and $T=0.1$. Note, the signal vanishes exactly at $X=1$.}
\label{fig:RIXSvsOmega_Wi}
\end{figure}

First, in Fig.~\ref{fig:RIXSvsOmega_Wi}, we plot the resonant part of the RIXS response as a function of the transfered energy, $\Omega$, for various 
transfered momenta, $X$, for three different incident photon energies, $\omega_{i}=-0.5,\; 0.5,\; 1.5$. For $\omega_{i}=-0.5$ the incident photon 
energy is lying in the lower Hubbard band and the inelastic relaxation processes can happen only within the lower Hubbard band which is evident in the 
single peak structure in the RIXS response in Fig.~\ref{fig:RIXSvsOmega_Wi}(a). At the $M$ point the peak is large and well defined but as we go towards 
the middle of the Brillouin zone ($X=0.0$) the peak gets broadened and the position of the peak does not disperse significantly. Finally, as 
we approach the zone center the position of the peak disperses significantly and moves towards lower energy, also at the same time the peak gets more 
and more well defined though the integrated spectral intensity under the peak gradually diminishes and eventually goes to zero at the center of the 
Brillouin zone ($X=1$) which is related to the vanishing of the uniform charge susceptibility.   

As we increase the incident photon energy above the Mott gap, we start to excite the system into the upper Hubbard band and in 
Fig.~\ref{fig:RIXSvsOmega_Wi}(b), we show a characteristic response when the incident photon energy $\omega_{i}=0.5$ is at the bottom of the upper Hubbard 
band. Near the zone corner the response still shows a single peak structure albeit shifted by the insulating gap, but as we go towards the zone center, the 
response at low energy develops a very narrow secondary peak separated from the broad main peak by the Mott gap. The low-energy peak arises due to 
relaxation processes within the upper Hubbard band and is non-dispersive in nature, whereas the high-energy peak arises due to relaxation processes across 
the Mott gap into the lower Hubbard band and is found to be dispersive over the Brillouin zone. With further increase in the incident photon energy to 
$\omega_{i}=1.5$ the low energy peak and the spectral weight under it, as shown in Fig.~\ref{fig:RIXSvsOmega_Wi}(c), grows significantly and is visible 
for all momenta along the $\langle 1 \cdots 1 \rangle$ direction. However the intensity of this low-nergy peak shows non-monotonic behaviour - it first 
increases up to $X=0$ and then starts to decrease and eventually vanishes at the zone center, $X=1$. On the other hand the intensity of the high-energy 
peak monotonically decreases as well as disperses to lower energies as we go from the zone corner towards the zone center and eventually vanishes at the 
center of the Brillouin zone. 
 
\begin{figure}
\includegraphics[scale=0.9,clip]{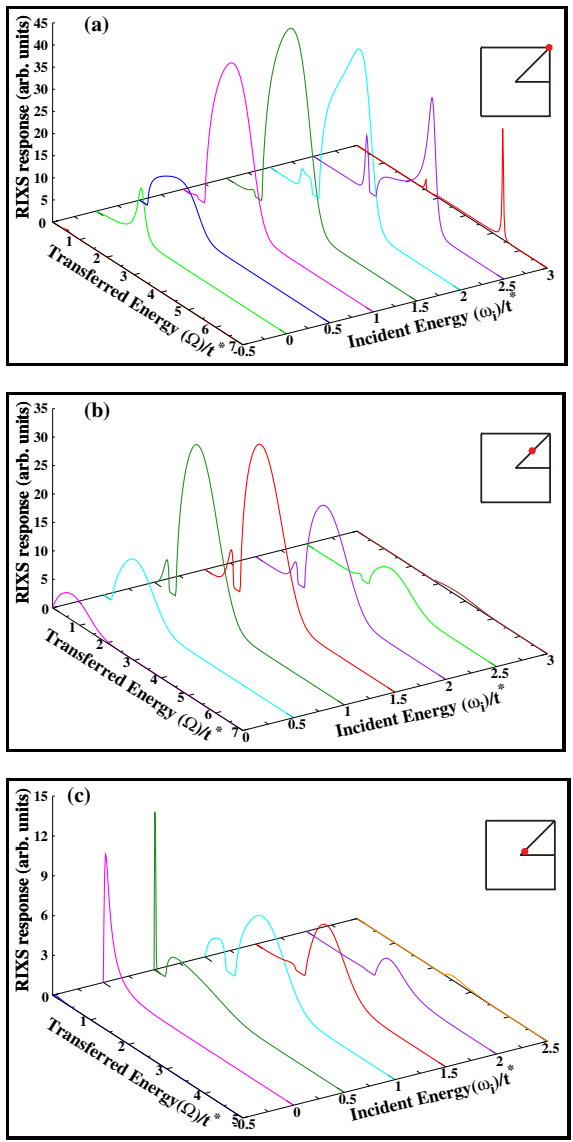}
\caption{(Color online) RIXS response as a function of transfered energy ($\Omega$) for various fixed incident photon energies, $\omega_{i}$, varying from 
the bottom of the lower Hubbard band to the top of the upper Hubbard band for three different momenta, $X$. Panel (a) zone corner ($M$ point) $X=-1$, 
(b) somewhere in the middle of the zone $X=0$ and (c) near the zone center $X=0.9$. Inset of each plot shows the position of each $X$ inside the first 
Brillouin zone. All other parameters are the same as in Fig.~\ref{fig:RIXSvsOmega_Wi}.}
\label{fig:RIXSvsOmega_X}
\end{figure}

In Figs.~\ref{fig:RIXSvsOmega_X}(a), (b) and (c), we show the systematic evolution of the RIXS response for three momenta $X = -1$, $X = 0$ and $X = 0.9$, 
respectively. As we vary the incident photon energy from the bottom of the lower Hubbard band to the top of the upper Hubbard band the RIXS response 
evolves from a single-peak structure to a two-peak structure and when $\omega_{i}$ is beyond the edges of the Hubbard bands the response vanishes quickly 
due to an exponential reduction of the density of states which in effect drastically reduces the phase space for inelastic scattering. Also, the overall 
response decreases as we go towards zone center.
 
\begin{figure}
\includegraphics[scale=0.3,clip]{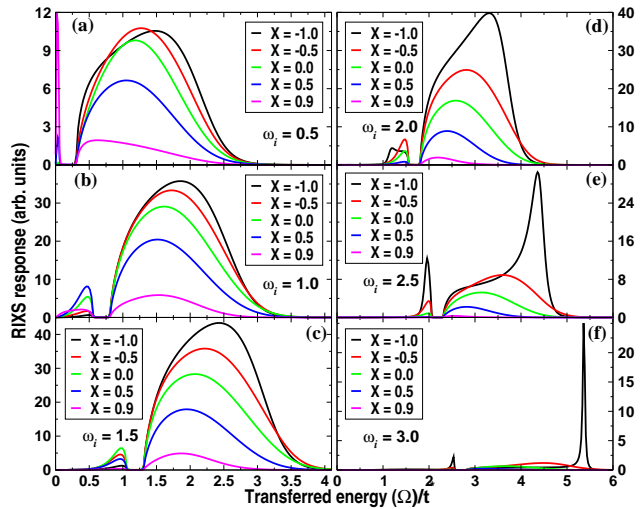}
\caption{(Color online) Systematic evolution of the RIXS response as a function of transfered energy ($\Omega$) for six incident photon energies from the 
bottom of the upper Hubbard band, panel (a) $\omega_{i} = 0.5$, to the top of the upper Hubbard band, panel (f) $\omega_{i}=3.0$. The dispersive nature of 
the high energy peak (Mott gap excitation) is noticeable. All other parameters are same as in Fig.~\ref{fig:RIXSvsOmega_Wi}.}
\label{fig:PanelPlotRIXSvsOmegaForWi}
\end{figure}

In Fig.~\ref{fig:PanelPlotRIXSvsOmegaForWi}, we show a more detailed and systematic evolution of the RIXS response with varying incident energy in 
the upper Hubbard band. As we can clearly see, the low-energy peak which arises due to relaxation processes within the upper Hubbard band does not disperse 
over the Brillouin zone (except its intensity varies) whereas the high-energy peak (which arises due to relaxation processes across the Mott gap) shows 
significant dispersion over the Brillouin zone. This feature has been observed in RIXS measurements on a Mott insulator 
$\mbox{Ca}_{2}\mbox{CuO}_{2}\mbox{Cl}_{2}$ by Hassan \textit{et. al.}~\cite{HasanScience2000} and has been attributed to strong correlation effects.    
\begin{figure}
\includegraphics[scale=0.9,clip]{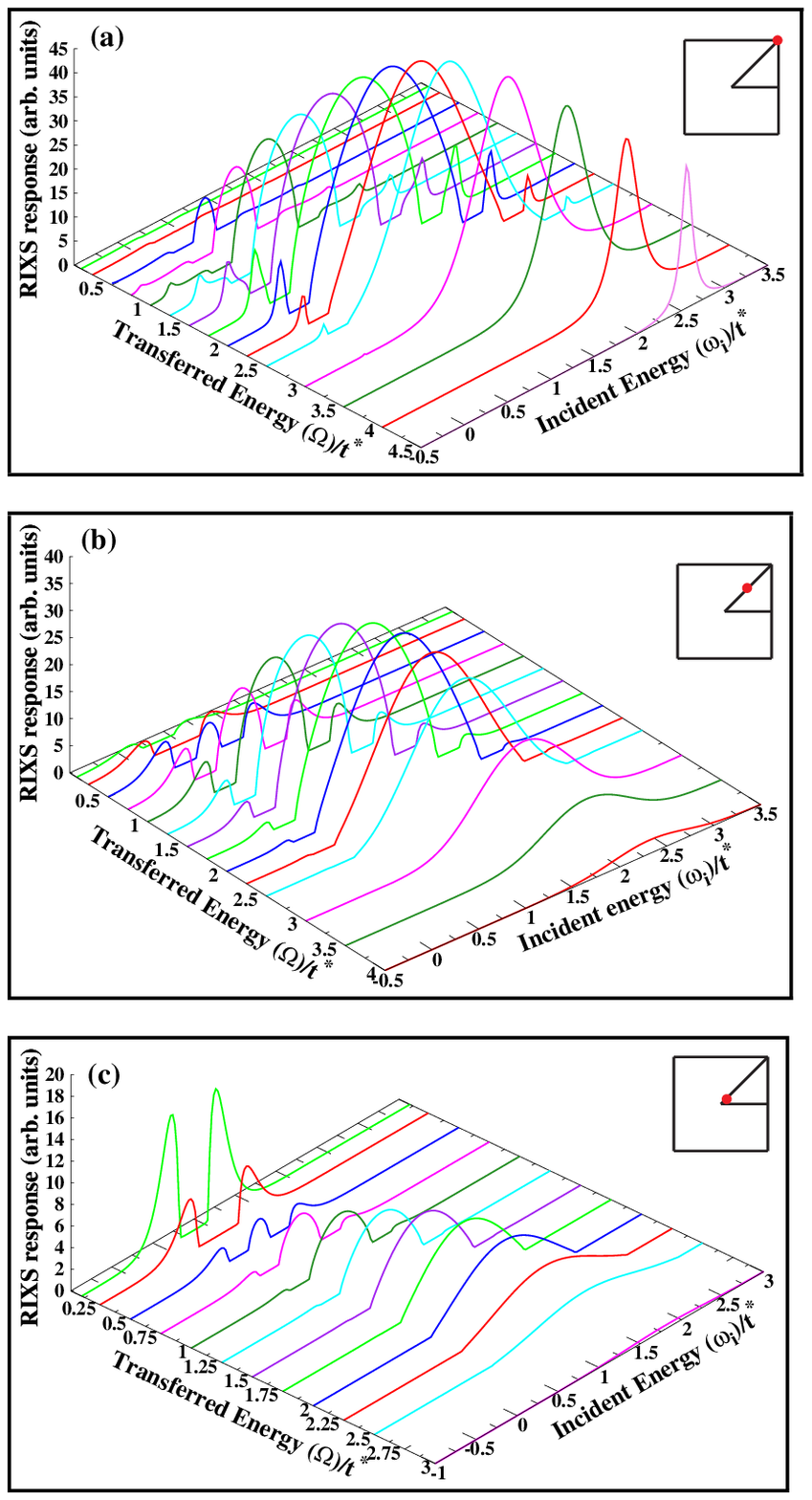}
\caption{(Color online) RIXS response as a function of incident energy, $\omega_{i}$, for various fixed transfered energies, $\Omega$, for three 
characteristic transfered momentum, $X$. Panel (a) zone corner ($M$ point) $X=-1$, (b) in the middle of the zone $X=0$ and (c) near the 
zone center $X=0.9$. The inset of each plot shows the position of each $X$ inside the first Brillouin zone. All other parameters are the same as in 
Fig.~\ref{fig:RIXSvsOmega_Wi}.}
\label{fig:RIXSvsWi_X}
\end{figure}

It is interesting to mention that at zero temperature in the half-filled Mott-insulating ground state, the chemical potential as well as the Fermi level 
lies within the Mott gap which results in a completely filled lower Hubbard band (LHB) and a completely empty upper Hubbard band (UHB). As a result, when 
the incident photon energy is within the LHB the RIXS response will vanish due to the unavailability of any unoccupied state in the LHB to which the core 
electron can be excited and when the incident photon energy is within the upper Hubbard band the RIXS response will have a single peak structure arising 
due to particle-hole excitations across the gap in the presence of the core-hole potential. At finite temperature some of the states near the top of the 
LHB thermally excite across the gap and occupy the bottom of the UHB. So, at finite temperature, if the incident photon energy is within the LHB, the core 
electron can be excited to thermally excited unoccupied states of the LHB, giving rise to a peak corresponding to the relaxation processes within the LHB 
and if the incident photon energy is in the UHB then the core electron that is excited to the empty states of the UHB can undergo relaxation processes 
either through scattering from the thermally excited UHB electrons occupying the bottom of the UHB giving rise to a low-energy nondispersive peak or 
through particle-hole excitations across the gap giving rise to a dispersive high-energy peak. Also, the whole structure gets shifted to higher energy with 
increasing incident photon energy due to the fact that with increasing incident photon energy the transfered energy must also increase in order to have 
resonant scattering from  the thermally excited states which predominantly occupy the bottom of the UHB and the top of the LHB. In the case of large gap 
Mott insulator, as will be shown in a following section (Sec. VI), the intraband relaxation processes from the thermally excited states are negligibly small 
compared to the excitations across the Mott gap.   
  
Finally, in Figs.~\ref{fig:RIXSvsWi_X}(a), (b) and (c), we plot the RIXS response as a function of the incident photon energy, $\omega_{i}$, for various 
fixed transfered photon energies, $\Omega$ for three transfered momenta $X=-1$, $X=0$ and $X=0.9$, respectively. For small $\Omega$, the RIXS response 
shows a two-peak structure in $\omega_{i}$ which corresponds to the relaxation processes within the individual bands (upper and lower Hubbard bands). As 
we increase $\Omega$ an additional peak develops between the two peaks. This peak corresponds to the interband relaxation processes across the Mott gap 
and grows very rapidly with increasing $\Omega$ while the other two peaks decrease in intensity until we are finally left with a lone peak. Hence, we can 
infer that in a Mott insulator, interband relaxation processes across the Mott gap are dominant over intraband relaxation processes. Also, as we go from 
the zone corner to the zone center the intensity of the peaks decreases as in Fig.~\ref{fig:RIXSvsOmega_Wi}.

\begin{figure}[htbp]
\includegraphics[scale=0.3,clip]{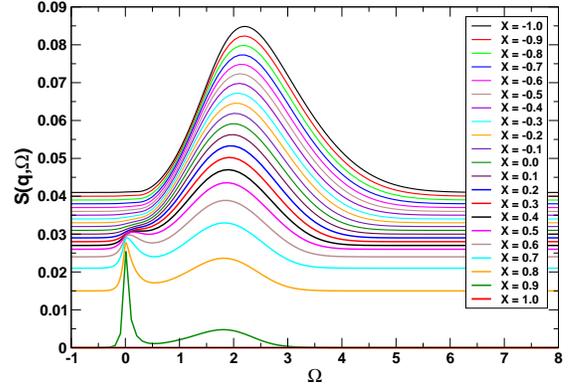}
\caption{(Color online) Dynamical structure factor, $S(\mathbf{q},\Omega)$ for U=2.0 and $T=0.1$. Note, it vanishes for $X=1$.}
\label{fig:S_q_w}
\end{figure}
We also have calculated the dynamical structure factor, $S(\mathbf{q},\Omega)$, which is proportional to the nonresonant part of the RIXS 
response~\cite{FreericksPRB2000}. In Fig.~\ref{fig:S_q_w}, we plot $S(\mathbf{q},\Omega)$ for the small-gap insulator at $T=0.1$. Near the zone corner, 
$S(\mathbf{q},\Omega)$ has a broad midinfrared peak but as we go towards the zone center, a secondary peak develops near $\Omega=0$ and the integrated 
spectral weight under the midinfrared peak decreases and finally exactly at the zone center ($X=1$) the midinfrared peak completely vanishes while the 
peak around $\Omega=0$ turns into a delta function which again arises due to the vanishing of the uniform charge susceptibility at finite frequency. 
This behavior is quite different from the resonant response where the two-peak structure is most prominent near the zone corner and also the midinfrared 
peak in $S(\mathbf{q},\Omega)$ is much smaller and much less dispersive than the high-energy peak observed in the resonant response and most importantly 
the position of the peak cannot be identified with any particular X-ray transition process.
\section{Core-hole broadening effects}
In the preceding section, we have not included any additional core-hole lifetime broadening effects which can arise due to various nonradiative Auger and 
fluorescence effects and are important in the transition metal RIXS processes (we only included the intrinsic many-body effects in determining the 
core-hole lifetime). In our calculation, we can easily include such effects by simply making the core-hole energy $E_{h}$ complex \textit{i. e.} by making 
the transformation $E_{h} \rightarrow E_{h}-i\Gamma$ into the \textit{retarded} Green's function, $G_{h}^{r}(t)= \Theta(t)\left[G_{h}^{>}(t)-G_{h}^{<}(t)
\right]$, and $E_{h} \rightarrow E_{h}+i\Gamma$ into the \textit{advanced} Green's function $G_{h}^{a}(t)= \Theta(-t)\left[G_{h}^{<}(t)-G_{h}^{>}(t)\right]$
, respectively. Here, $\Gamma = \frac{h}{\tau}$ with $\tau$ being the core-hole life time and for most materials $\Gamma \sim 100 - 400\mbox{ meV}$. 

\begin{figure}[htbp]
\includegraphics[scale=0.3,clip]{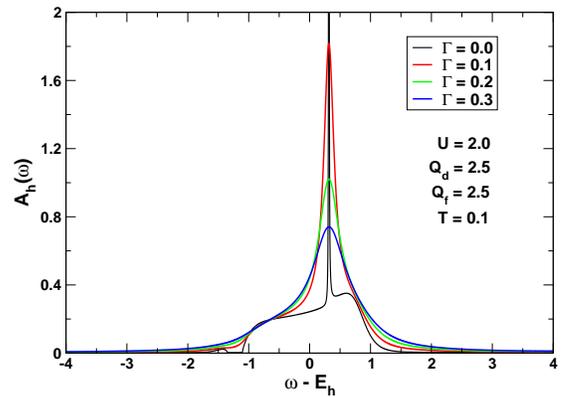}
\caption{(Color online) Core-hole spectral function, $A_{h}(\omega)$, evolution with core-hole hole broadening parameter, $\Gamma$. All other parameters 
are same as in Fig.~\ref{fig:RIXSvsOmega_Wi}.}
\label{fig:AhTotVsGamaU2}
\end{figure}
First, in Fig.~\ref{fig:AhTotVsGamaU2}, we show the systematic evolution of the core-hole spectral function, $A_{h}(\omega)$, with additional core-hole 
broadening effects parametrized by $\Gamma$. With increasing $\Gamma$, the height of the sharp peak in $A_{h}(\omega)$ reduces while its width 
significantly broadens but the asymmetrical nature of the structure is largely retained. Also, the Mott gap in the $\Gamma = 0.0$ case (near 
$\omega-E_{h}\sim -1.0$) is replaced by a dip in the spectral function and the tail of $A_{h}(\omega)$ increases with increasing $\Gamma$.  

\begin{figure}
\includegraphics[scale=0.32,clip]{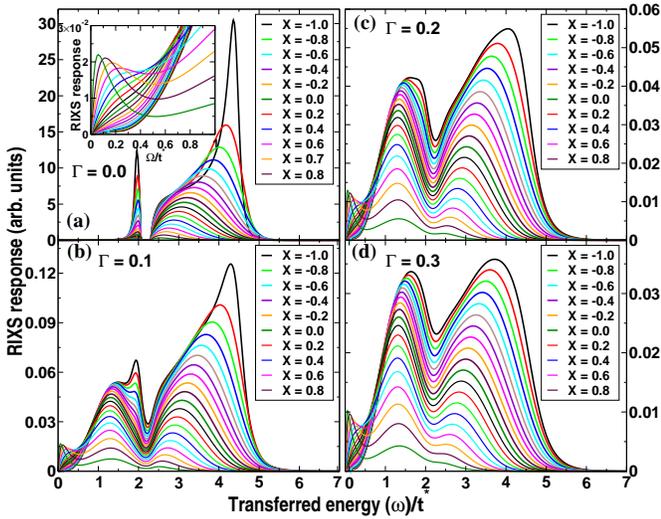}
\caption{(Color online) Evolution of RIXS response as a function of the transfered energy, $\Omega$, and various fixed transfered momenta, $X$, for 
incident photon energy $\omega_{i}=2.5$ with various core-hole lifetime broadening parameters, $\Gamma$. Panel (a), response for $\Gamma=0$ and the inset 
of panel (a) shows quasielastic peak in the blown up far-infrared region. Panels (b), (c) and (d) show responses for broadening parameters much smaller 
($\Gamma=0.1$), comparable ($\Gamma = 0.2$) and larger ($\Gamma=0.3$) than the Mott insulating gap $\Delta_{gap}\sim 0.25$. All other parameters are the 
same as in Fig.~\ref{fig:RIXSvsOmega_Wi}. For all the four sets, calculations have been performed from $X=-1.0$ (top most curve) to $X=0.9$ (bottom most 
curve) with a momentum interval $0.1$.}
\label{fig:RIXSvsGama}
\end{figure}
In Figs.~\ref{fig:RIXSvsGama} (a), (b), (c) and (d), we show a detailed evolution of the RIXS response with various broadening parameters, $\Gamma$ 
(measured in units of $t^{*}$). We choose four characteristic parameters $\Gamma=0.0$, $\Gamma=0.1$, $\Gamma=0.2$ and $\Gamma=0.3$ corresponding to no 
broadening, much smaller, comparable and larger broadening compared to the intrinsic Mott-insulating gap in the system, respectively. All other parameters 
chosen are the same as in the previous case. In Fig.~\ref{fig:RIXSvsGama} (a), we plot the RIXS response without any additional core-hole broadening effect
 and the response shows a clear two-peak structure with the peaks well separated by the Mott gap. The low-energy peak is nondispersive but the high-energy
peak is highly dispersive in momenta and also changes its shape significantly as we go from the zone corner to the zone center. Also, interestingly, close 
to the zone center a quasielastic peak develops in the far-infrared region. The intensity as well as sharpness of this peak increases as we go towards the 
zone center.

In the presence of additional core-hole lifetime broadening effects, we no longer see a clear gap structure in Figs.~\ref{fig:RIXSvsGama} (b), (c) and (d) 
but the two-peak structure is still clearly evident for all momenta. Also the high-energy peak still remains dispersive throughout the Brillouin zone, 
while the low-energy midinfrared peak is more or less nondispersive in nature as in the case with no additional core-hole broadening. The height of the 
quasielastic peak close to the zone center decreases with increasing $\Gamma$, which is related to the reduction of the overall RIXS response with 
decreasing core-hole lifetime, while its width and dispersive features remains similar to the $\Gamma=0.0$ case. However, in a real experiment, finite 
resolution and the resolution broadened tail of the huge elastic peak will mask such quasielastic features which will then not be observable. So, the 
presence of additional core-hole broadening effects can significantly modify the overall response, but the most important qualitative features, like the 
two-peak structure and the dispersive features of these peaks remain similar.
\section{Response away from half-filling}
\begin{figure}[htbp]
\includegraphics[scale=0.8,clip]{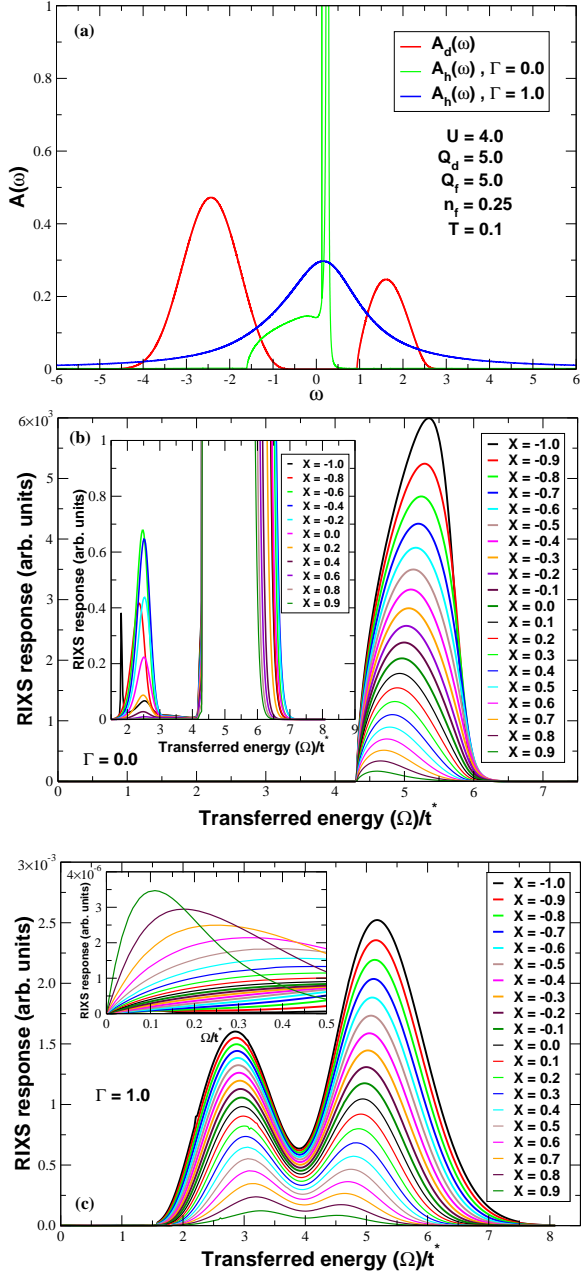}
\caption{(Color online) Panel (a) : $d$-hole spectral function, $A_{d}(\omega)$, as well as the core-hole spectral function, $A_{h}(\omega)$ with 
($\Gamma = 1.0$, blue line) and without ($\Gamma = 0.0$, green line) core-hole broadening effects in a particle-hole asymmetric large gap Mott insulator. 
For the core-hole spectral function, energy is measured with respect to the core-hole energy, $E_{h}$, \textit{i. e.} $\omega\rightarrow\omega-E_{h}$. 
Panels (b) and (c), RIXS response as a function of the transfered energy, $\Omega$, for various fixed transfered momenta, $X$, without [$\Gamma = 0.0$, 
panel (b)] and with [$\Gamma = 1.0$, panel (c)] core-hole broadening effects for incident photon energy $\omega_{i}=3.5$. Magnified response in the inset 
of panel (b) clearly shows a \textit{two-peak} structure and the inset of panel (c) shows the quasielastic peak in the blown up far-infrared region. The 
parameters used for this plot are $U=4.0$, $Q_{d}=Q_{f}=5.0$, $T=0.1$ and $n_{f}=0.25$.}
\label{fig:RIXSmultiplotU4}
\end{figure}

Finally, we consider a Mott insulator at arbitrary filling to examine the breaking of particle-hole symmetry in the RIXS response. We choose $U=4.0$, 
$Q_{d}=Q_{f}=5.0$, $n_{f}=0.25$ and $T=0.1$. This choice of parameters gives a large-gap Mott insulator ($\Delta_{gap}\sim\mbox{ 1.8}$ in units of $t^{*}$).
 In Fig.~\ref{fig:RIXSmultiplotU4}(a), we show the $d$-hole spectral function, $A_{d}(\omega)$, as well as the core-hole spectral function , 
$A_{h}(\omega)$, with ($\Gamma = 1.0$) and without ($\Gamma = 0.0$) core-hole broadening effects. $A_{d}(\omega)$ is dominated by two asymmetrical peaks 
separated by a large gap at the Fermi level. $A_{h}(\omega)$ [without any additional core-hole broadening effect ($\Gamma=0.0$)] shows two very closely 
spaced sharp peaks on top of a broad feature but with the inclusion of large core-hole broadening ($\Gamma = 1.0$) the whole structure gets drastically 
modified and $A_{h}(\omega)$ resembles a broad nearly symmetrical single peak. 

In Fig.~\ref{fig:RIXSmultiplotU4}(b), we show the RIXS response as a function of the transfered energy, $\Omega$, for various fixed transfered momenta, 
$X$, for $\Gamma = 0.0$. The response in this case is overwhelmingly dominated by a huge peak arising due to relaxation processes across the Mott gap into 
the LHB, while the intraband (within the UHB) relaxation processes as shown in the inset of Fig.~\ref{fig:RIXSmultiplotU4}(b) are much weaker than the 
interband processes. As already has been mentioned in a previous section (Sec. IV), in a large-gap Mott insulator the density of thermally excited states 
which occupy the bottom of the UHB is extremely small and hence cannot provide any significant relaxation to the core electrons excited to the UHB. The 
high-energy peak, just as in the case of particle-hole symmetric half-filled case, shows significant dispersion with transfered momentum - the peak 
disperses outwards in energy as we go from the zone center towards the zone corner. Finally, we study the RIXS response in the presence of finite 
core-hole broadening. In Fig.~\ref{fig:RIXSmultiplotU4}(c), we show results for a typical broadening (of the order of the Mott gap) $\Gamma=1.0$. The 
first noticeable feature is the reemergence of the \textit{two-peak} structure. This is mainly due to huge suppression of the sharp resonating peak in 
$A_{h}(\omega)$, as can be observed in Fig.~\ref{fig:RIXSmultiplotU4}(a), which in effect drastically reduces the resonating response across the Mott gap. 
The high-energy peak still shows significant dispersion across the entire Brillouin zone and the low-energy peak also shows dispersive features. Also, as 
shown in the inset of Fig.~\ref{fig:RIXSmultiplotU4}(c), a very weak quasielastic peak similar to the half-filled case emerges which is related to the 
long tail in the core-hole spectral function.
\section{Conclusions}
In conclusion, we have studied the RIXS response in a Mott insulator which is modelled by the Falicov-Kimball model. We have considered both the 
particle-hole symmetric half-filled case as well as the general particle-hole asymmetric case. We find that when the incident photon energy is lying 
within the upper Hubbard band, the resonant response shows a two-peak structure arising from the intraband (low-energy peak) and interband (high-energy 
peak) relaxation processes (as expected since the ``gap" is larger but the temperature is the same as before). The high-energy peak is found to be much 
larger and sometimes overwhelmingly larger (away from half-filling case) than the low-energy peak and shows dispersive features throughout the entire 
Brillouin zone, while the low-energy peak remains more or less nondispersive. These distinctive features have already been observed in a large class of 
transition metal K-edge RIXS responses in a wide class of oxide materials and have been attributed to the nonlocal nature of the Mott gap excitations.

We also have considered moderately large core-hole broadening effects on the RIXS response and we see that despite significant change in the RIXS response 
many interesting qualitative features like the two-peak structure and the dispersive nature of the high-energy peak remains more or less intact. The 
quasielastic feature near the zone center becomes comparable to the other two peaks. However, this peak will be completely masked by the resolution limited
tail of the elastic peak (and cannot be observed in any current experiments). For the half-filled case, we also have calculated the dynamical structure 
factor, $S(\mathbf{q},\Omega)$, which is proportional to the nonresonant part of the response. $S(\mathbf{q},\Omega)$ is either dominated by a very weakly 
dispersive midinfrared peak when the transfered photon momentum, $X$, is near the zone corner or by a narrow peak around $\Omega = 0$ when the momentum is 
close to the zone center and exactly at the zone center $S(\mathbf{q},\Omega)$ vanishes for finite frequencies, while the narrow peak becomes a delta 
function peak.\\ 

\section*{ACKNOWLEDGEMENTS}
This work was supported by the US Department of Energy under grant no. DE-FG02-08ER46542 for the work at Georgetown and grant no. DE-FG02-08ER46540 for 
the CMSN collaboration. We would like to thank Tom Devereaux and Brian Moritz for many useful discussions and a critical reading of this manuscript. We 
would also like to thank M. A. van Veenendaal, A. Bansil, R. Markiewicz, J. Moreno, Z. Hussain, J. Rehr and A. Sorini for useful discussions. JKF also 
acknowledges support from the McDevitt bequest at Georgetown. All the Feynman diagrams in this paper were drawn by using open source software program 
JaxoDraw and the original reference has been duly cited in Ref.~\onlinecite{Jaxodraw}.

\end{document}